\documentstyle[11pt,epsf]{article}

\title{\vspace{0cm}\large\bf
Cosmological Evolution with $\Lambda$-Term \\
and Any Linear Equation of State
}
\author{
Alexander S. Silbergleit\\
Gravity Probe B, W.W.Hansen Experimental Physics Laboratory,\\
Stanford University, Stanford, CA 940305-4085, USA\\
 e-mail: {\it gleit@relgyro.stanford.edu}
}

\date{~}

\begin{document}

\maketitle

\begin{abstract}
\noindent Recent observational indications of an accelerating universe enhance the interest in studying models with a cosmological constant. We investigate cosmological expansion (FRW metric) with $\Lambda>0$ for a general linear equation of state $p=w\rho$, $w>-1$, so that the interplay between cosmological vacuum and quintessence is allowed, as well.

Four closed-form solutions (flat universe with any $w$, and 
$w=1/3$, $-1/3,\, -2/3$) are given, of which the last one appears to be new. For the open universe a simple relation between solutions with different parameters is established: it turns out that a solution with some $w$ and (properly scaled) $\Lambda$ is expressed algebraically via another solution with special different values of these parameters.

The expansion becomes exponential at large times, and the amplitude at the exponent depends on the parameters. We study this dependence in detail, deriving various representations for the amplitude in terms of integrals and series. The closed-form solutions serve as benchmarks, and the solution transformation property noted above serves as a useful tool. Among the results obtained, one is that for the open universe with relatively small cosmological constant the amplitude is independent of the equation of state. Also, estimates of the cosmic age through the observable ratio $\Omega_\Lambda/\Omega_M$ and parameter $w$ are given; when inverted, they provide an estimate of $w$, i. e., the state equation, through the known age of the universe.


\end{abstract}

\vfill\eject

\section{Introduction}

Recent data on the brightness of distant SN Ia ([1], [2]; see also [3] and the references therein), as well as evidence coming from the cosmic age, large scale structure, and cosmic microwave background anisotropy combined with the cluster dynamics, 
indicate, most probably, that the observed cosmological expansion is  accelerating. Perhaps the most natural, although definitely not unique, reason for this acceleration is the presence of a cosmic vacuum of nonzero energy and pressure; because of that, investigation of various cosmological models including a positive cosmological constant becomes interesting, once again.
In this paper we study one such model involving matter with an arbitrary (linear) equation of state; in particular, an interplay between vacuum and quintessence [4] is studied.

We consider a Friedmann cosmology described by the Friedmann-Robertson-Walker metric ,
\begin{equation}\label{metr}
ds^2= - dt^2 + a^2(t)\left( \frac{dr^2}{1-kr^2} + r^2d\Omega^2\right) \; ,
\end{equation}
where $d\Omega$ is the element of solid angle and $k= -1, 0$, or $1$, according to whether the universe is open, flat, or closed (we use the units with $G=c=1$). The only energy-momentum tensor consistent with homogeneity and isotropy of the universe is the one corresponding to a perfect fluid, described by the energy density $\rho$ and  pressure $p$. The latter are assumed to satisfy the equation of state
\begin{equation}\label{eqst}
 p=w\rho,\qquad w={\rm const}>-1\; .
\end{equation}
The range of values of $w$ is chosen on the grounds that $w=-1$ corresponds to the vacuum equation of state, and vacuum contributions are already included in our model with the cosmological constant $\Lambda>0$. Moreover, no physical reasons are known so far to justify the values of $w$ which are more negative that the vacuum one, $w=-1$; it is not easy to imagine the kind of physics that would lead to matter with the pressure larger than its density. But the following analysis does not depend at all on whether $w$ is larger or smaller than one, so we do not limit this parameter from above. Condition (\ref{eqst}) allows for quintessence , that is, for $-1<w<0$; investigation of its influence on cosmological expansion is one of the goals of this paper.

Under these assumptions, Einstein's equations reduce to
\begin{equation}\label{goveq}
3\left(\frac{1}{a} \frac{da}{dt}\right)^2= 8 \pi  \rho + \Lambda - 3\, \frac{k}{a^2},\qquad
\frac{d\rho}{dt} = -3\,\frac{\rho + p}{a}\,\frac{da}{dt}\; ,
\end{equation}
the second one expressing energy conservation. By (\ref{eqst}), this equation is easily integrated giving the density in terms of the scale factor,
\begin{equation}\label{rho}
\rho = \frac{M}{a^{3(w+1)}}\; ,
\end{equation}
where the constant $M>0$ characterizes the abundance of matter in the universe. A substituion of (\ref{rho}) in the first of equations (\ref{goveq}) produces the equation for the scale factor $a(t)$ only,
\begin{equation}\label{sfeq}
\left(\frac{da}{dt}\right)^2= \frac{8\pi M}{3}\,\frac{1}{a^{3w+1}}  + 
\frac{\Lambda}{3}\,a^2 - k
\; ;
\end{equation}
we append it with the Big Bang initial condition 
\begin{equation}\label{initcon}
a(0)=0\; .
\end{equation}
Note that the initial value of the density, $\rho(0)$, is infinite for $w>-1/3$, and zero for $-1<w<-1/3$; it is finite and positive for $w=-1/3$, which value corresponds to the so called Einstein's quintessence\footnote{The term `Einstein quintessence' was coined recently in [5] on the grounds of the effective equation of state for Einstein's cosmological solution of 1917.}.

Equation (\ref{sfeq}) with the initial condition (\ref{initcon}) constitute the problem studied below; its solution, together with formula (\ref{rho}), specifies completely the corresponding cosmological expansion at all times.

\section{Large--Time Expression for the Scale Factor}
\subsection{Problem Reformulation and Parameter Range}

For convenience, we rescale the variables according to
\begin{equation}\label{var} 
t=\sqrt{\frac{3}{\Lambda}}\,\tau,\qquad a(t)=\left(\frac{8\pi M}{\Lambda}\right)^{1/3(w+1)}y(\tau)\; ,
\end{equation}
and introduce new parameters:
\begin{equation}\label{par} 
 \nu\equiv3w+1,\quad \nu>-2;\quad \omega\equiv\frac{3}{\Lambda}\,
\left(\frac{\Lambda}{8\pi M}\right)^{2/3(w+1)}=
3\left(8\pi M\right)^{-2/(\nu+2)}\left(\Lambda\right)^{-\nu/(\nu+2)}>0\; .
\end{equation}
Denoting by a dot the derivative with respect to $\tau$, we rewrite (\ref{sfeq}) and (\ref{initcon}) as
\begin{equation}\label{ydot} 
\dot{y}^2=y^{-\nu}+y^2-k\omega,\qquad y(0)=0;\qquad \nu>-2;
\qquad k=0,\pm1;\quad \omega>0\; ;
\end{equation}
since we are studying the expansion, a positive square root of the right hand side of (\ref{ydot}) is, of course, always taken for $\dot{y}$.

The equation and initial condition (\ref{ydot}) are consistent for any value of parameter $k\omega$ if $\nu$ is positive, because at small times $\dot{y}^2\approx y^{-\nu}>0$. For $\nu=0$ at small times one gets $\dot{y}^2\approx1-k\omega$, so $\omega$ is limited by unity for the closed universe ($k=1$). When $-2<\nu<0$, the governing equation in the beginning of the expansion becomes $\dot{y}^2\approx-k\omega$, so only open and flat universe ($k=-1,\,0$) solutions are possible.

Moreover, we are interested only in the solutions describing an unbounded cosmological expansion, i. e., those with $y(\tau)\to\infty$ at $\tau\to\infty$. A simple mechanical analogy helps to establish parameter limits ensuring that. Namely, equation (\ref{ydot}) describes the motion of a particle of a unit mass along the positive $y$ axis, starting at the origin, in the potential
\[
V(y)=-\left(y^{-\nu}+y^2\right)\leq0\; ;
\]
the total energy of the particle is $-k\omega$. When $-2<\nu<0$, the potential is monotonically decreasing, with the largest value $V(0)=0$. Hence any such motion with a non-negative total energy ($k=-1,0$, open and flat universe) is infinite. The same monotonicity of the potential holds for $\nu=0$, but the largest value now is $V(0)=-1$, which yields the already known restriction $\omega<1$ for any infinite motion in the case $k=1$ (closed universe). 

Finally, for $\nu>0$ the potential is strictly negative and non-monotonic: at $y=y_\nu>0$ such that $V^{'}(y_\nu)=0$ it reaches a unique maximum $V(y_\nu)=-\omega_\nu$ given by
\begin{equation} \label{omega_nu}
 \omega_\nu\equiv\min_{y\geq0}\,\left(y^{-\nu}+y^2\right)=y_\nu^{-\nu}+y_\nu^2
=\frac{\nu+2}{\nu}\,y_\nu^2,\qquad y_\nu=\left(\frac{\nu}{2}\right)^{1/(\nu+2)}\; .
\end{equation}
An infinite motion requires the total energy to be larger than that maximum, $-k\omega>-\omega_\nu$, which is automatically fulfilled for the open and flat universe; however, for the closed universe ($k=1$), the parameter $\omega$ proves to be limited from above, $\omega<\omega_\nu$. This constraint has a clear physical meaning: to overwhelm the curvature effect in the closed universe and get an unbounded expansion, one has to have enough vacuum repulsion, that is, a large enough $\Lambda$. But for $\nu>0$ the parameter $\omega$ is proportional to a negative power of $\Lambda$ [see (\ref{par})], hence $\omega$ cannot be too large.

Note that $y_\nu\to+0$ and $\omega_\nu\to1+0$ when $\nu\to+0$, which allows us to  describe the cases of positive and zero $\nu$ uniformly. Also, both $y_\nu$ and  $\omega_\nu$ tend to unity from above when $\nu\to\infty$, hence they have at least one maximum. In fact, each of the functions, plotted in Fig.1, has one maximum exactly: $\max \omega_\nu=2$ at $\nu=2$ (for some reason, it is relativistic gas that has the largest `store of expansion' in the closed universe), and $\max y_\nu\approx1.149$ at $\nu\approx7.182$. Some values of particular interest are: for $\nu=0$ ($w=-1/3$,  Einstein's quintessence), $y_\nu=0$, $\omega_\nu=1$;  for $\nu=1$ ($w=0$, pressureless `dust'), $y_\nu=2^{-1/3}\approx0.794$, $\omega_\nu=3\times2^{-2/3}\approx1.890$; for $\nu=2$ ($w=1/3$, relativistic gas), $y_\nu=1$, $\omega_\nu=2$; for $\nu=4$ ($w=1$, Zeldovich's superstiff fluid), $y_\nu=2^{1/6}\approx1.122$, and again $\omega_\nu=3\times2^{-2/3}\approx1.890$. 

Summarizing, the range of parameters of interest is
\begin{equation} \label{range}
 k=1:\quad \nu\geq0,\;0<\omega<\omega_\nu; \qquad
k=0,-1:\quad \nu>-2,\; 0<\omega<\infty\; .
\end{equation}

\subsection{Expansion Amplitude and Its Basic Properties}

With parameters in the range (\ref{range}), the cosmological expansion goes indefinitely [unique solution to problem (\ref{ydot}) grows unboundedly], and at large times it is exponential in time, since for $y\gg\max\{1,\,\omega\}$ the equation becomes $\dot{y}\approx y$ [for $k=0,\,1$ the large time condition is apparently relaxed to just $y\gg1$]. Therefore
\begin{equation} \label{A_nu}
y(\tau)\equiv y(\tau,\,\nu,\,k\omega)=A_\nu(k\omega)\,e^\tau\,[1+o(1)]\;,\qquad \tau\to\infty\; ,
\end{equation}
and we are particularly interested in the behavior of the expansion amplitude $A_\nu(k\omega)$ for the whole parameter range (\ref{range}). For $\nu\geq0$ and $k\omega=\omega_\nu$ (i. e., $k=1,\,\,\omega=\omega_\nu$) the solution $y(\tau)$ reaches the (unstable) rest point $y_\nu$ in an infinite time, so $A_\nu(\omega_\nu)$ should be zero, which remains to be proved.

We derive an explicit expression for the amplitude by integrating equation (\ref{ydot}), 
\begin{equation} \label{tauy}
\tau(y)\equiv \tau(y,\,\nu,\,k\omega)=\int_0^y\frac{dx}{\sqrt{x^{-\nu}+x^2-k\omega}}\; .
\end{equation}
For $y,\tau\to\infty$ the integral here diverges logarithmically, as it should. To get the desired expression for the amplitude, one needs to extract this logarithm of $y$, that is, to add and subtract something in the integrand allowing for a closed--form quadrature. The most natural choice seems the same integrand with $k=0$, in which case one writes: 
\[
\tau(y)=\frac{2}{\nu+2}\,\ln\left[y^{(\nu+2)/2}+\sqrt{1+y^{(\nu+2)}}\right]+
\int^y_{0}\left[\frac{1}{\sqrt{x^{-\nu}+x^2-k\omega}}-\frac{1}{\sqrt{x^{-\nu}+x^2}}\right]\,dx=\;
\] 
\begin{equation} \label{tauas}
\ln\left[2^{2/(\nu+2)}\,y\right]+
\int^\infty_{0}\left[\frac{1}{\sqrt{x^{-\nu}+x^2-k\omega}}-\frac{1}{\sqrt{x^{-\nu}+x^2}}\right]\,dx+o(1),\qquad y\to\infty\;
\end{equation}
[unlike (\ref{tauy}), the integral here converges at infinity]. This expression can be combined with the equality (\ref{A_nu}) to write the desired fomula for the amplitude as
\begin{equation} \label{fi}
A_\nu(k\omega)=A_\nu(0)\,e^{-\phi_\nu(k\omega)},\qquad 
\phi_\nu(k\omega)\equiv\int^\infty_{0}\left[\frac{1}{\sqrt{x^{-\nu}+x^2-k\omega}}-\frac{1}{\sqrt{x^{-\nu}+x^2}}\right]\,dx \; ,
\end{equation}
where
\[
A_\nu(0)=2^{-2/(\nu+2)}\; 
\]
(note that, naturally, $\phi_\nu(0)=0$).

Two features of the expansion amplitude behavior can be established immediately. First, when $k\omega=\omega\to \omega_\nu-0$, the integral (\ref{fi}) for $\phi_\nu(k\omega)$ diverges at  $x=y_\nu$, so $\phi_\nu(k\omega)$ goes to $+\infty$; therefore $A_\nu(\omega_\nu)=0$, as expected. 

Next, we have, for a given $\nu$:
\begin{equation} \label{derA}
\frac{d \ln A_\nu(k\omega)}{d(k\omega)}=-\,\frac{d\phi_\nu(k\omega)}{d(k\omega)}\; ,
\end{equation}
and the latter derivative is readily found from (\ref{fi}):
\begin{equation} \label{derfi}
\frac{d\phi_\nu(k\omega)}{d(k\omega)}=\frac{1}{2}\,\int_0^{\infty}
\frac{dx}{\left(x^{-\nu}+x^2-k\omega\right)^{3/2}}>0\; .
\end{equation}
Therefore $A_\nu(k\omega)$ is a decreasing function of $k\omega$ for any given relevant $\nu$; in particular,
\begin{equation} \label{ofc}
A_\nu(-\omega)>A_\nu(0)[>A_\nu(\omega),\; \nu\geq0,\;\omega<\omega_\nu]\; . 
\end{equation}
That means that, as one expects intuitively, for the same parameters $w$, $M$, and $\Lambda$, the amplitude for the open universe is larger than that for the flat universe which, in its turn, is larger than the amplitude for the closed universe if the latter exists. Generally, by figuring out the signs of the curvature and parameter $\nu$ and taking into account expression (\ref{par}) for $\omega$, one can see that, given $k=\pm1,\;\nu$, and the matter abundance $M$, the growth amplitude is larger for larger cosmological  constant $\Lambda$. That clearly makes physical sense: the larger the contribution of the vacuum repulsion, the stronger the expansion is.

\section{Exact Solutions: Flat Universe, Relativistic Gas,\\ Einstein Quintessence, Quintessence with $w=-2/3$,\\ and Others}

Exact and, especially, elementary solutions are always most valuable; in the context of this study, they specifically shed light on how the expansion amplitude behaves.

It is not difficult to see that all the integrals involved in the above solution prove to be elementary in the following four cases.

\subsection{Flat Universe: $k=0$}

For $k=0$ equation (\ref{tauy}) gives the result presented in part in the previous section; here is how it looks in full: 
\begin{equation} \label{y(0)}
\tau(y)=\frac{2}{\nu+2}\,\ln\left[y^{(\nu+2)/2}+\sqrt{1+y^{(\nu+2)}}\right]\;,
\qquad  y(\tau)=\left[\sinh\left(\frac{\nu+2}{2}\,\tau\right)\right]^{2/(\nu+2)}\; .
\end{equation}
Thus for flat universe the amplitude of the exponential growth of the scale factor is
\begin{equation} \label{A(0)}
A_\nu(0)=2^{-2/(\nu+2)},\qquad 0<A_\nu(0)<1\; .
\end{equation}

Note also the small time asymptotics expression, 
\[
y(\tau)=\left(\frac{\nu+2}{2}\,\tau\right)^{2/(\nu+2)}[1+o(1)],\qquad \tau\to+0\; ,
\]
which is valid for all the values of $k\omega$ and $\nu>0$ (the curvature and cosmological constant do not play any role in the early universe). For $\nu=0$, and $\nu<0,\,\,k=-1,$ the expansion becomes linear with the time, in the beginning. 

\subsection{Relativistic Gas: $w=1/3$}

Here $\nu=2$, and we have:
\[
\tau(y)=\frac{1}{2}\,
\ln\frac{2y^2-k\omega+2\sqrt{y^{4}-k\omega y^2+1}}{2-k\omega}\;,
\]
\begin{equation} \label{y2}
 y(\tau)=
\sqrt{
\frac{2-k\omega}{4}\exp(2\tau)-\frac{2+k\omega}{4}\exp(-2\tau)+\frac{k\omega}{2}
}=
\sqrt{
\sinh(2\tau)-k\omega\,\sinh^2\tau
}
\; ,
\end{equation}
so that
\begin{equation} \label{A2}
A_2(k\omega)=\frac{\sqrt{2-k\omega}}{2}\;  
\end{equation}
(all these results agree with the above ones for $k=0$). Note that $\omega_2=2$ [the definition of $\omega_\nu$ is given in (\ref{omega_nu}) ], and the expression (\ref{A2}) for the amplitude is well defined exactly up to $\omega=\omega_2=2$, for the closed universe ($k=1$).

\subsection{Einstein quintessence: $w=-1/3$}

In this case we have $\nu=0$, and
\begin{equation} \label{nu0}
\tau(y)=\ln\frac{y+\sqrt{1-k\omega+y^{2}}}{\sqrt{1-k\omega}}\;,
\qquad  
y(\tau)=\sqrt{1-k\omega}\,\sinh\,\tau\; ,
\end{equation}
hence
\begin{equation} \label{A0}
A_0(k\omega)=\frac{\sqrt{1-k\omega}}{2}\; ,
\end{equation}
again in agreement with section 3.1. Of course, as seen from (\ref{omega_nu}), $\omega_0=1$.

\subsection{Quintessence with $w=-2/3$}

Finally, here $\nu=-1$, thus
\[
\tau(y)=\ln\frac{2y+2\sqrt{y^{2}+y-k\omega}}{2\sqrt{-k\omega}+1},
\qquad  
\qquad k=0,-1\;;
\]
\begin{equation} \label{y-1}
y(\tau)=\frac{1+2\sqrt{-k\omega}}{4}\exp(\tau)+\frac{1}{2}\,\left[\exp(-\tau)-1\right]=
\sqrt{-k\omega}\,\sinh\tau+\sinh^2(\tau/2)
\; ,
\end{equation}
and
\begin{equation} \label{A-1}
A_{-1}(k\omega)=\frac{1+2\sqrt{-k\omega}}{4}\; ,
\end{equation}
Naturally, these expressions remain physically meaningful only for $k=-1,\,0$, coinciding with the results of section 3.1 in the latter case of flat universe. 
Note that solutions (\ref{y(0)}), (\ref{y2}) and (\ref{nu0}) are known (see [6] and the references therein), although perhaps not always in exactly the same form; however, solution (\ref{y-1}) appears to be new.

\subsection{Exact Solutions in Terms of Elliptic Functions}

By definition, an elliptic integral is the integral of any rational function of the integration variable and the square root of a polynomial of this variable whose degree is three or four. Any elliptic integral can be reduced to a combination of elementary functions and three standarad elliptic integrals (see c. f. [7]).

The form of integrals (\ref{tauy}) and (\ref{fi}) determining our cosmological solution and its large-time asymptotics prompts an arbitrary power change of the integration variable to try reducing them to elliptic integrals. Such an attempt turns out to be successful for four different positive values of parameter $\nu$. By the transformation property of section 5, four reciprocal negative values should be added to the list, making a total of eight different exact solutions in terms of elliptic functions. Since these representations are rather cumbersome and not very informative, we choose not to give them here explicitly; instead, in the table below we show the corresponding values of $\nu$ and $w$, and the proper substitution making the integrals elliptic. 
\vskip3mm
\centerline{\bf Elliptic function solutions for $\nu>0$}
\vskip2mm
\centerline{
\begin{tabular}{|c|c|c|}\hline
 $\nu$&$w$ & Substitution\\ \hline\hline
6&5/3&$x=\xi^{1/2}$\\ \hline
4&1&$x=\xi^{-1/2}$\\ \hline
1&0&$x=\xi^{-1}$\\ \hline
2/3&-1/9&$x=\xi^{2}$\\ \hline
\end{tabular}}
\vskip3mm

Note that $w=0$ describes pressureless matter (`dust'), and $w=1$ corresponds to Zeldovich's superstiff fluid [8]. According to (\ref{transol}), the reciprocal negative values of $\nu$ allowing for the elliptic function solutions are $\nu=-3/2\,-4/3,\,-2/3,\,-1/2$, $k=-1$.

\section{Estimates of the Scale Factor and Expansion Amplitude}

In the general case, no closed--form solution is available, so it is useful to have at least some estimates illustrating the behavior of cosmological expansion. Here we derive two groups of such estimates differing by the parameters involved in them. It is not difficult to check that all the exact elementary solutions satisfy these estimates, as they must.

\subsection{Estimates Independent of the State Equation}

Some estimates of the scale factor and the expansion amplitude follow immediately from (\ref{ydot}). Indeed, we have
\[
\dot{y}=\sqrt{y^{-\nu}+y^2-k\omega}>\sqrt{y^2-k\omega}
\]
for all $y\geq0$ when $k=-1,\,0$, and for $y\geq\sqrt{\omega}$ when $k=1$. From here we  obtain:
\begin{equation}\label{est}
\frac{d}{d\tau}\left(\sqrt{y^2-k\omega}\right)\,>\,1,\qquad y\geq0,\quad k=-1;\qquad y\geq\sqrt{\omega},\quad k=1\; .
\end{equation}
In the case of an open universe we integrate this between zero and an arbitrary moment of time, which, due to the initial condition (\ref{ydot}), results in
\[
y+\sqrt{y^2+\omega}\,>\,\sqrt{\omega}\,\exp\tau\; .
\]
Solving this inequality for $y$ provides a simple lower bound of the open universe expansion  at all times:
\begin{equation}\label{estyo}
y(\tau)=y(\tau,\nu,-\omega)\,>\,\sqrt{\omega}\,\sinh\tau
,\qquad \tau\geq0\;\;({\rm resp.,}\,\,y\geq0),\quad ,\quad k=-1;\; .
\end{equation}
Sending $\tau\to\infty$ and comparing to (\ref{A_nu}), we find the estimate for the growth amplitude:
\begin{equation}\label{estAo}
A_\nu(-\omega)\,\geq\,\sqrt{\omega}/2
,\qquad  k=-1\; .
\end{equation}

In the case of closed universe we integrate inequality (\ref{est}) up to an arbitrary moment of time starting with [see (\ref{tauy})]
\begin{equation}\label{tauom}
\tau_\omega\equiv\tau(\sqrt{\omega},\nu,\omega)=\int_0^{\sqrt{\omega}}\frac{dx}{\sqrt{x^{-\nu}+x^2-\omega}}\; ,
\end{equation}
with $y$ changing, respectively, from $\sqrt{\omega}$ to $y(\tau)$. Thus we arrive at
\[
y+\sqrt{y^2-\omega}\,>\,\sqrt{\omega}\,\exp(\tau-\tau_\omega)\; ;
\]
solution of this inequality yields the estimate
\begin{equation}\label{estyc}
y(\tau)=y(\tau,\nu,\omega)\,>\,\sqrt{\omega}\,\cosh(\tau-\tau_\omega)
,\qquad \tau\geq\tau_\omega\;\;({\rm resp.,}\,\,y\geq\sqrt{\omega}),\quad k=1\; .
\end{equation}
Again, in the limit $\tau\to\infty$ this inequality combined with (\ref{A_nu}) allows for the corresponding estimate of the expansion amplitude:
\begin{equation}\label{estAc}
A_\nu(\omega)\,\geq\,\left(\sqrt{\omega}/2\right)\,\exp(-\tau_\omega)
,\qquad  k=1,\;\nu\geq0\; .
\end{equation}
Note that, by (\ref{tauom}), $\tau_\omega\to\infty$ when $\omega\to\omega_\nu-0$, hence the estimate (\ref{estAc}) does not contradict the amplitude approaching zero at $\omega_\nu$. Clearly, all these estimates do not explicitly depend on $\nu$, except the dependence on it of $\tau_\omega$ in the case of a closed universe.

\subsection{Estimates Independent of Parameter $\omega$}

On the physical grounds, one expects that the time in which the expansion reaches a given value of the scale factor, $y$, increases from open to flat to closed universe,
\[
\tau(y,\nu,-\omega)\,<\,\tau(y,\nu,0)\,[<\,\tau(y,\nu,\omega),\quad \nu\geq0]\; ,
\]
because of the curvature effect. Formally, this is easily derived from the general solution (\ref{tauy}) by dropping the term $k\omega$ in the denominator of the integrand there, and choosing the inequality sign accordingly. Using the exact solution (\ref{y(0)}) for the flat universe, we thus obtain:
\[
 \tau<\frac{2}{\nu+2}\,\ln\left[y^{(\nu+2)/2}+\sqrt{1+y^{(\nu+2)}}\right],\qquad k=-1,\;\nu>-2;
\]
\begin{equation}\label{tauflat}
\end{equation}
\[
\tau>\frac{2}{\nu+2}\,\ln\left[y^{(\nu+2)/2}+\sqrt{1+y^{(\nu+2)}}\right],\qquad k=1, \;\nu\geq0
 \; .
\]
By inverting these, we get the estimates for the scale factor:
\[
 y(\tau)>\left[\sinh\left(\frac{\nu+2}{2}\,\tau\right)\right]^{2/(\nu+2)},\qquad k=-1,\;\nu>-2;
\]
\begin{equation}\label{yflat}
\end{equation}
\[
y(\tau)<\left[\sinh\left(\frac{\nu+2}{2}\,\tau\right)\right]^{2/(\nu+2)}
,\qquad k=1, \;\nu\geq0
 \; .
\]
Naturally, the estimates for the growth amplitude implied by (\ref{yflat}) are nothing else than (\ref{ofc}), in view of (\ref{A(0)}). 

\section{Transformation Property of Cosmological Solutions  and Expansion Amplitude for the Open Universe}

\subsection{Derivation of Transformation Property}

A family of cosmological solutions describing an open universe turns out to have a useful transformation property. It relates, in a rather simple algebraic way, a solution with a given set of parameters $\nu,\,\,\omega$ to another with different values of these parameters. 

To establish this property, let us consider the initial value problem (\ref{ydot}) with $k=-1$, $k\omega=-\omega$, and introduce a new unknown function  $z=z(\tau)=z(\tau, \nu, -\omega)$ instead of $y=y(\tau)=y(\tau, \nu, -\omega)$ according to
\begin{equation} \label{sub}
y=\omega^{1/2} \,z^{2/(\nu+2)};\qquad 
\dot{y}=\frac{2\omega^{1/2}}{\nu+2}\,z^{-\nu/(\nu+2)}\,\dot{z}\; .
\end{equation}
Since $\nu+2>0$, the initial condition for $z$ is the same as for $y$,
\begin{equation} \label{init}
z(0)=0\; .
\end{equation}
On the other hand, equation (\ref{ydot}) after substitution (\ref{sub}) becomes
\[
\left(\frac{2}{\nu+2}\right)^2\,\dot{z}^2=
\omega^{-\nu/2-1}+ z^{(4+2\nu)/(\nu+2)}+z^{-\left(-2\nu/(\nu+2)\right)}\; .
\]
Introducing the new time,
\begin{equation} \label{tau}
\tilde{\tau}=\frac{\nu+2}{2}\,\tau\; , 
\end{equation}
and denoting by a prime the derivative, we rewrite the last equation as
\begin{equation} \label{eveq}
\left({z}^{'}\right)^2=z^{-\left(-2\nu/(\nu+2)\right)}+ z^{2}+\omega^{-(\nu+2)/2}\; .
\end{equation}

Clearly, the initial value problem (\ref{eveq}), (\ref{init}) coincides with the original problem (\ref{ydot}) in which $\tau$ is replaced with $\tilde{\tau}$, $\nu$ with 
$-2\nu/(\nu+2)$, and $\omega$ with $\omega^{-(\nu+2)/2}$, so that, by the uniqueness theorem,
\[
z=z(\tilde{\tau},\, \nu,\,-\omega)=y\left((\nu+2)\tau/2,\,-2\nu/(\nu+2),\,-\omega^{-(\nu+2)/2}\right)\; .
\]
But then, according to (\ref{sub}) and (\ref{tau})
\begin{equation} \label{transol}
y(\tau,\, \nu,\, -\omega)\equiv\omega^{1/2}
\left[
y\left((\nu+2)\tau/2,\,-2\nu/(\nu+2),\,-\omega^{-(\nu+2)/2}\right)
\right]^{2/(\nu+2)},\qquad k=-1\; ,
\end{equation}
where the identity sign stresses that the equality holds for all moments of time, as well as for any $\nu>-2$ and $\omega>0$. Sending $\tau$ to infinity provides, by (\ref{A_nu}), the corresponding transformation property of the expansion amplitude,
\begin{equation} \label{tranamp}
A_{\nu}(-\omega)\equiv\omega^{1/2}\,
\left[
A_{-2\nu/(\nu+2)}\left(-\omega^{-(\nu+2)/2}\right)
\right]^{2/(\nu+2)},\qquad k=-1\; .
\end{equation}
In particular, our previous results (\ref{nu0}) and (\ref{A0}) for $\nu=0$ satisfy (\ref{transol}) and (\ref{tranamp}), respectively. 

Note also the corresponding transformation of cosmic ages as functions of the scale factor $y$, 
\begin{equation} \label{transage}
\tau(y,\,\nu,\,-\omega)=\frac{2\omega^{1/2}}{\nu+2}\,
\tau
\left(
\left(y/\omega^{1/2}\right)^{(\nu+2)/2},\,-2\nu/(\nu+2),\,-\omega^{-(\nu+2)/2}
\right)\; ,
\end{equation}
which is easily obtained either by directly inverting (\ref{transol}), or by properly changing the integration variable in (\ref{tauy}).

\subsection{Parameter Duality Map}

The correspondence of the indeces involved in (\ref{transol}),
\[
{\cal M}_\nu\,:\,\nu\longrightarrow -\frac{2\nu}{\nu+2},
\]
is a one--to--one map of the semiaxis $\nu\in(-2,\,\,\infty)$ into itself with the only fixed point $\nu=0$, so that $(0,\,\,\infty)$ is mapped into $(-2,\,\,0)$, and vice versa. Therefore for the case of the open universe it is enough to study cosmological solutions with either positive or negative $\nu$, getting then the results for the opposite sign of this parameter immediately from (\ref{transol}) and (\ref{tranamp}); we are intensively using this approach below. 

Moreover, it is easy to see that ${\cal M_\nu}$ is a duality map, that is, ${\cal M_\nu}={\cal M_\nu}^{-1}$, or ${\cal M_\nu}^2={\cal I}$, ${\cal I}$ being the unit map. The whole range of values of $\nu$ is thus represented as a set of pairs of values reciprocal under the map ${\cal M_\nu}$, with $\nu=0$ reciprocal to itself. In particular, one reciprocal pair is $\nu=2$ (relativistic gas) and $\nu=-1$; it is straightforward to see that $A_{2}(k\omega)$ from (\ref{A2}) and $A_{-1}(k\omega)$ from (\ref{A-1}) satisfy (\ref{tranamp}). 

Note that for the physical parameter $w$ entering the equation of state, the corresponding duality relation is
\[
{{\cal M}_w}\,:\,w\longrightarrow -\frac{w+5/9}{w+1},\qquad w>-1\; .
\]
It gives a one-to-one map of the interval $(-1,\;-1/3)$ into the semiaxis $(-1/3,\;\infty)$ and vice versa; its only fixed point $w=-1/3$ corresponds to Einstein's quintessence, which once again separates one group of cases from the other. The reason for that is well known, and it is seen from the equation for the universal acceleration, which is obtained by the obvious manipulation with the governing equations (\ref{goveq}):
\begin{equation}\label{accel}
\frac{3}{a} \frac{d^2a}{dt^2}=-4 \pi (\rho+3p) + \Lambda= -4 \pi (3w+1)\rho + \Lambda\; .
\end{equation}
Therefore matter with $w<-1/3$ enhances the accelerating effect of vacuum, while matter with $w>-1/3$ diminishes it, and it is only Einstein's quintessence, which is thus said to have no gravitational mass, that does not contribute to the acceleration. So, given the matter abundance $M$ and $k=-1$, an expansion with $w>-1/3$ and certain $\Lambda$ can in principle be related only to some expansion with $w<-1/3$ and smaller $\Lambda$, as it occurs according to (\ref{transol}).  Of course, the Einstein's quintessence solution cannot reciprocate with any other, having a different equation of state.

\subsection{Open Universe with $\omega\gg1$}

As a first example of using the transformation property, let us consider the asymptotic behavior of the expansion amplitude for the open universe ($k=-1$) when $\omega$ is large, which is easily found from (\ref{tranamp}) and (\ref{A(0)}). Indeed, in this case $\omega^{-(\nu+2)/2}\to0$, and we have: 
\begin{equation} \label{bigom}
A_\nu(-\omega)=\omega^{1/2}\,
\left[
A_{-2\nu/(\nu+2)}\left(0\right)
\right]^{2/(\nu+2)}\left[1+o(1)\right]=
\frac{\omega^{1/2}}{2}\left[1+o(1)\right]\; .
\end{equation}
This result agrees with all the exact solutions (\ref{A2}), (\ref{A0}), and (\ref{A-1}), as well as with the estimate (\ref{estAo}). 

Thus the main term giving the large--time behavior of the open universe with $\omega\gg1$  does not depend on the equation of state. The corresponding full asymptotic series, which proves to be converging, is found in section 6.5; all higher order corrections involved in it do depend on $\nu$, i. e., on the state equation.

\section{Expansion Amplitude at $\omega$ Near the Cutoff Value} 

As explained in section 2, cosmological expansion remains infinite, and its growth amplitude is thus well defined, up to $\omega=\omega_\nu$ for $\nu\geq0$ and closed  universe ($k=1$), which value forms thus a physical cutoff, in this case. For $-2<\nu<0$ and open universe the solution exists for all positive values of $k\omega$, so $k\omega=0$ should be also treated as a cutoff. The expansion amplitude behavior near the cutoff value, which cannot be smooth, is of interest in both cases.

We first consider $\nu>0$, $k=1$, and $\omega$ near  $\omega_\nu$, that is, $\omega_\nu-\omega\to+0$. In this case the main contribution to the integral (\ref{fi}) for $\phi_\nu(k\omega)$ comes from the vicinity of the point $x=y_\nu$, where
\begin{equation} \label{appr}
x^{-\nu}+x^2-k\omega\approx \omega_\nu-\omega+(\nu+2)(x-y_\nu)^2\; .
\end{equation}
Using this approximation it is not difficult to find that 
\[
\phi_\nu(\omega)= -\frac{1}{\sqrt{\nu+2}}\,\ln(\omega_\nu-\omega)+O(1)\; ,\qquad k=1,
\quad\omega\to \omega_\nu-0\;,
\]
and 
\begin{equation} \label{Acut}
A_\nu(\omega)=A_\nu(0)K_\nu\left(\omega_\nu-\omega\right)^{1/\sqrt{\nu+2}}[1+o(1)]\;,\qquad 
k=1,\quad\omega\to \omega_\nu-0\; .
\end{equation}
Calculation of the value of the constant $K_\nu$ requires more tedious analysis carried out in section 6.4, which provides
\begin{equation} \label{Knu}
K_\nu={\omega_\nu^{-1/\sqrt{\nu+2}}}\,{\exp\left[\sum_{n=1}^{\infty}\,C_n(\nu)\right]}\; ,
\end{equation}
where the coefficients $C_n(\nu)=O(1/n^2),\,\,n\to\infty$, are given explicitly in  (\ref{Cn}).

For Einstein quintessence, $\nu=0$, and any curvature, approximation (\ref{appr}) works uniformly in $\nu\geq0$ only for the function and its first derivative, but not for the second one. Therefore, with $\nu$ formally set to zero, it gives a wrong coefficient at the quadratic term (2 instead of 1). So for $\nu=0$ the right answer is given not by (\ref{appr}) with $\nu=0$, but by the exact formula (\ref{A0}) (rederived, by the way, in a different manner in section 7.3).

Finally, when $-2<\nu<0$, we should consider the open universe only, $k=-1$, in which case we can write, for $\omega\to+0$:
\[
\phi_\nu(k\omega)=\int^\infty_{0}\left[\frac{1}{\sqrt{x^{-\nu}+x^2+\omega}}-\frac{1}{\sqrt{x^{-\nu}+x^2}}\right]\,dx= 
\]
\[
\omega^{-1/2\nu}\int^\infty_{0}
\left[
\frac{1}{\sqrt{\xi^{-\nu}+\omega^{-(2+\nu)/\nu}\xi^2+1}}-
\frac{1}{\sqrt{\xi^{-\nu}+\omega^{-(2+\nu)/\nu}\xi^2}}
\right]\,d\xi=
O(\omega^{-1/2\nu})\; 
\]
[recall that $-1/2\nu,\,\,-(2+\nu)/\nu>0$]. So, by (\ref{fi}) and (\ref{A(0)}),
\begin{equation} \label{Asmalom}
A_\nu(-\omega)=A_\nu(0)\left[1+O(\omega^{-1/2\nu})\right],\quad 
A_\nu(0)=2^{-2/(\nu+2)},\qquad 
\omega\to+0\; ,
\end{equation}
The exact result (\ref{A-1}) for $A_{-1}(k\omega)$ agrees with this. Note that the remainder term in (\ref{Asmalom}) diverges at $\omega\to+0$, hence so does the derivative of $A_\nu(-\omega)$, as expected. Thus, in fact, the behavior of the growth amplitude at both cutoffs,  $\omega=\omega_\nu,\, \nu\geq0$, and $\omega=0,\,\nu<0$, is similar, with the only difference that the limit value of the amplitude is zero in the first case and positive in the second.

\section{Power series in  $k\omega$ for $A_\nu(k\omega)$ and its implications}

\subsection{Power series in  $k\omega$ for $\nu\geq0$}

 Repeated differentiation of the equality (\ref{derA}) in $k\omega$ provides
\begin{equation} \label{nderA}
\left[\,\ln A_\nu(k\omega)\,\right]^{(n)}
=-\phi^{(n)}_\nu(k\omega),\quad n=1,2,...;\qquad
f^{(n)}(k\omega)\equiv\,\frac{d^nf(k\omega)}{d(k\omega)^n}\; .
\end{equation}
Also, repeated differentiation of (\ref{derfi}) shows that, for $\nu\geq0$,
\begin{equation} \label{nderfi}
\phi^{(n)}_\nu(0)=\left(\frac{1}{2}\right)_n\,\int_0^{\infty}
\frac{dx}{\left(x^{-\nu}+x^2\right)^{n+1/2}},\qquad n=1,2,...;
\end{equation}
where the Pochhammer symbol $(\alpha)_n$ is defined in a usual way, $(\alpha)_0\equiv1,\,\,(\alpha)_n\equiv$\break$\Gamma(\alpha+n)/\Gamma(\alpha)=
(\alpha+n-1)(\alpha+n-2)\dots(\alpha+1)\alpha,\,\,n=1,2,...$, and $\Gamma(\alpha)$ is the Euler gamma--function. [Note that the integral in (\ref{nderfi}) diverges at the lower limit when $\nu<0$]. Therefore
\begin{equation} \label{Aser1}
\ln\left[\frac{ A_\nu(k\omega)}{A_\nu(0)}\right]
=-\sum_{n=1}^{\infty}\,\frac{\phi^{(n)}_\nu(0)}{n!}\,(k\omega)^n\; ,
\end{equation}
and the coefficients of this series are found to be (see Appendix)
\begin{equation} \label{Asercoe}
\frac{\phi^{(n)}_\nu(0)}{n!}=\frac{1}{2\sqrt\pi}\,
\frac{
\Gamma\left(\frac{2n}{\nu+2}+1\right)
\Gamma\left(\frac{\nu n}{\nu+2}+\frac{1}{2}\right)
}
{n},\qquad n=1,2,...\,.
\end{equation}
The radius of convergence and other properties of this series are studied below. Since the radius proves to be non-zero, function $A_\nu(k\omega)
$ is regular at $k\omega=0$. That means that, for a given positive $\nu$, the expansion amplitude for the open universe is an analytical continuation of that for the closed one, and vice versa.

\subsection{Case $\nu=2$}

The result (\ref{A2}) for $\nu=2$ is easily obtained from the series (\ref{Aser1}). Indeed, by (\ref{Asercoe}) we have
\[
\frac{\phi^{(n)}_2(0)}{n!}=\frac{1}{2\sqrt\pi}\,
\frac{
\Gamma\left(\frac{n}{2}+1\right)
\Gamma\left(\frac{n}{2}+\frac{1}{2}\right)
}{n\,\Gamma\left({n}+1\right)}=
\frac{1}{2\sqrt\pi}\,
\frac{\sqrt\pi\,\Gamma\left({n}+1\right)}{n\,2^n\,\Gamma\left({n}+1\right)}=
\frac{1}{n\,2^{n+1}}\; ,
\]
where we have used the double argument formula for the gamma-fucntion,
$2^{2z-1}\Gamma(z)\Gamma\left(z+1/2\right)=
\sqrt{\pi}\,\Gamma(2z)$, with $z=(n+1)/2$. Thus
\[
\ln\left[\frac{ A_0(k\omega)}{A_0(0)}\right]
=-\frac{1}{2}\,\sum_{n=1}^{\infty}\,\frac{(k\omega/2)^n}{n}=\frac{1}{2}\,\ln(1-k\omega/2)\;,
\]
and, according to (\ref{A(0)}), it is exactly the formula (\ref{A0}):
$A_2(k\omega)=0.5(2-k\omega)^{1/2}$. Note that the radius of convergence of the series (\ref{Aser1}) coincides with the cutoff value for this case, $2=\omega_2$. 

\subsection{Case $\nu=0$}

In this case formula  (\ref{Asercoe}) gives immediately $\phi^{(n)}_\nu(0)=1/2n$, so
\[
\ln\left[\frac{ A_0(k\omega)}{A_0(0)}\right]
=-\frac{1}{2}\,\sum_{n=1}^{\infty}\,\frac{(k\omega)^n}{n}=\frac{1}{2}\,\ln(1-k\omega)\;,
\]
that is, by (\ref{A(0)}), $A_0(k\omega)=0.5(1-k\omega)^{1/2}$ ,
which is exactly the expression (\ref{A0}). Here again the radius of convergence of series (\ref{Aser1}) coincides with the cutoff, $1=\omega_0$, hinting that this should probably be a generic property.

\subsection{General case $\nu>0$}

The general result for $\nu>0$ comes from studying the behavior of the coefficients of the series (\ref{Aser1}) with large numbers $n$. As demonstrated in the Appendix,
\begin{equation} \label{coefas}
\frac{\phi^{(n)}_\nu(0)}{n!}=\frac{1}{\sqrt{\nu+2}}\,
\frac{1}{n}\,\left(\frac{1}{\omega_\nu}\right)^n
\left[1+O\left(\frac{1}{n}\right)\right],
\qquad n\to\infty\,.
\end{equation}
This means that the convergence radius of the series (\ref{Aser1}) always coincides, as one would now expect, with the cutoff value $\omega_\nu$, and allows us to rewrite the series as
\[
\ln\left[\frac{ A_\nu(k\omega)}{A_\nu(0)}\right]
=-\frac{1}{\sqrt{\nu+2}}\,\sum_{n=1}^{\infty}\,\frac{(k\omega/\omega_\nu)^n}{n}+\,
\sum_{n=1}^{\infty}\,C_n(\nu)\,(k\omega/\omega_\nu)^n=
\]
\begin{equation} \label{Aser2}
\frac{1}{\sqrt{\nu+2}}\,\ln(1-k\omega/\omega_\nu)+\,
\sum_{n=1}^{\infty}\,C_n(\nu)\,(k\omega/\omega_\nu)^n\; ,
\end{equation}
where, in view of (\ref{Asercoe}) and (\ref{coefas}),
\begin{equation} \label{Cn}
C_{n}(\nu)=\frac{1}{\sqrt{\nu+2}}\,\frac{1}{n}
\left[
1-\frac{\omega_\nu^n}{2}\,\sqrt{\frac{\nu+2}{\pi}}\, 
\frac{
\Gamma\left(\frac{2n}{\nu+2}+1\right)
\Gamma\left(\frac{\nu n}{\nu+2}+\frac{1}{2}\right)
}{n!}
\right]=
O\left(\frac{1}{n^2}\right),
\qquad n\to\infty
\;.
\end{equation}
Expression (\ref{Aser2}) is useful for computing $A_\nu(k\omega)$ in the whole range $k\omega\leq \omega_\nu$. Also, for $k=1$ it determines completely the asymtotics of the amplitude near the cutoff $\omega_\nu$, whose main term is given above in (\ref{Acut}), (\ref{Knu}).

\subsection{Series in inverse powers of $\omega$ for $-2<\nu<0,\,\,k=-1$}

To obatin some series for $A_\nu(k\omega)$ with negative $\nu$ (and thus $k=-1$), we combine the above result (\ref{Aser2}) with the amplitude transformation property (\ref{tranamp}). Writing for brevity ${\mu}=-2\nu/(\nu+2)$, the latter provides
\[
\ln A_\nu(-\omega)=\ln\omega^{1/2}+\frac{1}{\sqrt{\nu+2}}\ln A_{{\mu}}(-\omega^{-\nu/(\nu+2)})=
\ln\frac{\omega^{1/2}}{2}+\frac{1}{\sqrt{\nu+2}}
\ln \left[\frac{A_{{\mu}}(-\omega^{-\nu/(\nu+2)})}{A_{{\mu}}(0)}\right]\; ,
\]
where we used the value of $A_{{\mu}}(0)$ from (\ref{A(0)}). Since, by definition, $\mu>0$, we can now replace the last term by its series expansion (\ref{Aser2}) provided that $\omega^{-\nu/(\nu+2)}<\omega_\mu$, or 
\begin{equation} \label{convcond}
\omega>-\frac{\nu+2}{\nu}\,\left(-\frac{\nu}{2}\right)^{2/(\nu+2)}
\end{equation}
[recall $(\nu+2),\,\,(-\nu)>0$]. Under this condition we thus obtain the desired series for $-2<\nu<0,\,\,k=-1$ in the form [the series coefficients are given in (\ref{Cn})]:
\begin{equation} \label{Aser3}
\ln A_\nu(-\omega)= \ln\frac{\omega^{1/2}}{2}+
\frac{1}{\sqrt{\nu+2}}\ln\left[1+\kappa_\nu(\omega)\right]+
\sum_{n=1}^{\infty}\,C_n(\mu)\,\left[\kappa_\nu (\omega)\right]^n\; ,
\end{equation}
\[
 {\mu}=-\frac{2\nu}{\nu+2},\qquad
0<\kappa_\nu(\omega)\equiv\frac{-\nu/2}{\left[-\nu\omega/\left(\nu+2\right)\right]^{(\nu+2)/2}}<1\; .
\]
This series converges if condition (\ref{convcond}) holds, and also provides the asymptotic formula
\begin{equation} \label{aslargom}
A_\nu(-\omega)= \frac{\omega^{1/2}}{2}
\left[1+ O\left(-\omega^{(\nu+2)/2}\right)\right],\qquad
\omega\to\infty,\quad -2<\nu<0,\quad k=-1 \; ,
\end{equation}
which sharpens expression (\ref{bigom}) by giving the exact order of the remainder.

\section{Discussion}

We have determined various significant properties of the cosmological solution, particularly, those of the expansion amplitude, and provided different expressions for the latter. The plots of $ A_\nu(k\omega)$ as a function of  $k\omega$ for several `popular' values of the parameter $\nu$, including positive, zero, and negative ones, are given in Fig. 2. 

An interesting question is what information on the universe can be gotten if the value of the expansion amplitude, say, $A_*$, is known? Even if no other information is available, but $A_*>1$, from (\ref{A(0)}) and the monotonicity of the amplitude established in sec. 2, one immediately concludes that the universe is open, since  $A_\nu(k\omega)<1$ for $k=1$ (and, of course, $\nu\geq0$). If, on the other hand, $0<A_*<1$, but the equation of state, i. e., parameter $\nu$, is more or less known, then equation 
\[
A_\nu(k\omega)=A_*
\]
provides, again due to the monotonic property, a unique solution for $k\omega$, which means the sign of the curvature, and a relation between $M$ and $\Lambda$ [see (\ref{par})].

One way of inferring the amplitude from observational data is as follows. Formula (\ref{rho}) for the density combined with (\ref{var}) reduces to
$8\pi\rho=\Lambda/y^{\nu+2}$, therefore
\[
\frac{\Omega_M}{\Omega_\Lambda}=\frac{8\pi\rho}{\Lambda}=\frac{1}{y^{\nu+2}}\; ,
\]
so that
\begin{equation} \label{y}
y=\left(\frac{\Omega_\Lambda}{\Omega_M}\right)^{1/(\nu+2)}\; .
\end{equation}
This equality holds at all moments of time. If, however, the expansion is in the asymptotic regime, (\ref{y}) and (\ref{A_nu}) allow for
\begin{equation} \label{Aobs}
A_\nu(k\omega)\approx 
\left(\frac{\Omega_\Lambda}{\Omega_M}\right)^{1/(\nu+2)}\exp(-\tau),\qquad
y(\tau)\gg 1\; .
\end{equation}
So, if the matter equation of state, the ratio $\Omega_\Lambda/\Omega_M$, and the properly scaled [see (\ref{var})] cosmic age are known, the expansion amplitude is known as well, if the asymptotic stage of the expansion is reached. By (\ref{y}), the latter requires
\begin{equation} \label{asreg}
\left(\frac{\Omega_\Lambda}{\Omega_M}\right)^{1/(\nu+2)}\gg1\; ,
\end{equation}
which, for $\nu$ not much smaller than negative unity, means just $\Omega_\Lambda/\Omega_M\gg1$. But in case the dominant form of matter in the universe is quitessence with the equation of state close to the vacuum one, condition (\ref{asreg}) might hold, and the expansion be in the asymptotic regime, despite the fact that $\Omega_\Lambda/\Omega_M>1$ is still of the order of unity.

One can also use (\ref{Aobs}) and (\ref{y}) to estimate the cosmic age. If (\ref{asreg}) is true [and moreover $\left(\Omega_\Lambda/\Omega_M\right)^{1/(\nu+2)}\gg\omega$ for open universe], that is, the large-time regime takes place,
(\ref{Aobs}) provides
\[
\tau\approx\frac{1}{\nu+2}\,\ln\frac{\Omega_\Lambda}{\Omega_M}-\ln A_\nu(k\omega)\; .
\]
Replacing the expansion amplitude with that of flat universe, $A_\nu(0)$, by (\ref{ofc}) and (\ref{A(0)}) leads to
\begin{equation} \label{ageaso}
\tau<\frac{1}{\nu+2}\,\ln\left(4\,\frac{\Omega_\Lambda}{\Omega_M}\right),\quad k=-1;\qquad \left(\frac{\Omega_\Lambda}{\Omega_M}\right)^{1/(\nu+2)}\gg\max\{1,\,\omega\}\; ,
\end{equation}
for open, and
\begin{equation} \label{ageasc}
\tau>\frac{1}{\nu+2}\,\ln\left(4\,\frac{\Omega_\Lambda}{\Omega_M}\right),\quad k=1;\qquad \left(\frac{\Omega_\Lambda}{\Omega_M}\right)^{1/(\nu+2)}\gg1\; 
\end{equation}
for closed universe in the asymptotic regime.

For any moment of the expansion the estimates for the age of the universe come from combining expression (\ref{y}) for the scale factor via observables with `geometrical' inequalities (\ref{tauflat}):
\begin{equation} \label{ages}
\tau\,<\,\tau_{Fl}(\Omega_\Lambda/\Omega_M,\,\nu),\quad k=-1;\qquad 
\tau\,>\,\tau_{Fl}(\Omega_\Lambda/\Omega_M,\,\nu),\quad k=1\; .
\end{equation}
Here
\begin{equation} \label{taufl}
\tau_{Fl}(\Omega_\Lambda/\Omega_M,\,\nu)\equiv
\frac{2}{\nu+2}\,\ln\left[\left(\frac{\Omega_\Lambda}{\Omega_M}\right)^{1/2}+\left(1+\frac{\Omega_\Lambda}{\Omega_M}\right)^{1/2}\right]
\end{equation}
is the exact age of the flat universe with the same parameters. If $\nu$, $\Omega_\Lambda/\Omega_M$, and the cosmic age are known accurately enough, its comparison with the bound (\ref{taufl}) allows for an immediate determination of the curvature of the universe. If, instead, only the last two parameters and the sign of the curvature are known, the state equation parameter can be estimated.

We now apply the above results to the currently favored cosmological model. It includes baryonic and dark matter, both with the same equation of state $w=0$ and total $\Omega_M=0.3\pm0.1$, and vacuum with $\Omega_\Lambda=0.7\pm0.1$. The third component is relativistic particles ($w=1/3$), but its abundance $\Omega_R<0.001$ is so small compared to the first two that it should be neglected. This results in exactly the studied model with $\nu=1$ and $\Omega_\Lambda/\Omega_M=1.5\div4$. Thus
\begin{equation} \label{Om/Om}
\left(\frac{\Omega_\Lambda}{\Omega_M}\right)^{1/(\nu+2)}=\left(\frac{\Omega_\Lambda}{\Omega_M}\right)^{1/3}=1.1\div 1.6 \sim 1\; ,
\end{equation}
so, by (\ref{y}), we are still rather far from the large time regime. (However,this is true for $\nu=1$ or any other $\nu$ not close to $-2$; otherwise the scale factor $y$ could be large enough. The main reason to assume the pressureless equation of state for dominating dark matter is that it is clumped around galaxies, which can happen only if its pressure is much smaller than its density. This is clearly true for non-negative pressure, $w\geq0$; but negative pressure of quintessence with the positive density might provide the clumping as well, a question which would be interesting to examine. In case the answer is positive, a better determination of the equation of state might become an important observational target). 

According to (\ref{taufl}) and (\ref{Om/Om}), for our universe
\begin{equation} \label{tauflour}
\tau_{Fl}=0.7\div1.0\; .
\end{equation}
The dimensional time is related to $\tau$ by [see (\ref{var})]
\begin{equation} \label{time}
T=\tau\,\sqrt{\frac{3}{G\Lambda}}=\tau\,(0.7\div0.8)\times 10^{18}\,sec=
\tau\,(21.3\div25.4)\,Gyr\; ,
\end{equation}
with
\[
\Lambda=8\pi\rho_V=8\pi(0.7\pm0.1)\rho_c=(1.0\div1.6)\times 10^{-28}\,g/cm^3\;
\]
based on
\[
\rho_c=(0.7\pm0.08)\times 10^{-29}\,g/cm^3\;,
\]
which corresponds to the range of the Hubble constant $H=70\pm10\,km/sec\cdot Mpc$. By virtue of (\ref{tauflour}) and (\ref{time}), for our universe
\begin{equation} \label{timefl}
T_{Fl}=\tau_{Fl}\,(21.3\div25.4)\,Gyr=(14.9\div17.8)\,Gyr\; ,
\end{equation}
that is, its age in case the universe is flat. Globular cluster data independently give the lower bound of the cosmic age as
\begin{equation} \label{ageobs}
T_u\geq(12\div16)\,Gyr\; . 
\end{equation}
According to these data, our universe is open if $T_u=(12\div14.9-0)\,Gyr$, and it could be either open or closed if $T_u=(14.9\div16)\,Gyr$. Since the first is true not only for the centers of both ranges (\ref{ageobs}) and (\ref{timefl}), but even for the right end of the first and the center of the second one, the conclusion is that our universe is most probably open.

Under this assumption we can now obtain a lower bound (\ref{ages}) of the cosmic age, to check the consistency of our estimates. Indeed, using (\ref{estyo}), we obtain an estimate of the parameter $\omega$ for our universe as
\begin{equation} \label{omour}
\omega\,<\,\frac{y^2}{\sinh^2\tau_u}=\frac{\left({\Omega_\Lambda}/{\Omega_M}\right)^{2/3}}{\sinh^2\tau_u}< 2.7\equiv\omega_{est}\;,
\end{equation}
with $\tau_u\approx1.54$ taken from the middle of the ranges (\ref{ageobs}) and (\ref{time}). This and the basic expression (\ref{tauy}) for the solution immediately produce
\[
\tau\,>\,
\int_0^{\left({\Omega_\Lambda}/{\Omega_M}\right)^{1/3}}
\frac{dx}{\sqrt{x^{-1}+x^2+\omega_{est}}}>
\int_0^{1.1}
\frac{dx}{\sqrt{x^{-1}+x^2+2.7}}\approx0.45
\; ,
\]
or, by (\ref{time}),
\[
T_u\,>\,(9.6\div11.4)\,Gyr\; ,
\]
in nice agreement with (\ref{ageobs}). 

Finally, to check the consistency of the current cosmological data, we can convert the estimate (\ref{ages}) for an open universe into an upper bound for the state equation parameter $\nu$, namely:
\begin{equation} \label{estnuo}
\nu\,<\,2\,\left\{\tau_u^{-1}
\ln\left[\left(\frac{\Omega_\Lambda}{\Omega_M}\right)^{1/2}+\left(1+\frac{\Omega_\Lambda}{\Omega_M}\right)^{1/2}\right]-1\right\}\; .
\end{equation}
By (\ref{ageobs}), (\ref{time}) and (\ref{Om/Om}), for our universe this provides
\[
\nu\,<\,0.5\div10\; .
\]
This is almost entirely consistent with the choosen $\nu=1$ (pressureless dust), except for a small range of parameters, when the ratio $\Omega_\Lambda$ is close to its minimum observed value $0.6$ and simultaneously the cosmic age is close to its maximum of $16\,Gyr$.

In conclusion, we point out two possible generalizations of the present study. The first of them deals with the phase transition, that is, with a sudden change of the equation of state, when $w$ changes abruptly at some moment of time. This case, as well as the one with a whole sequence of phase transitions, can be investigated in a fashion similar to the above.

The second way to generalize the study is to consider multicomponent matter, when
\[
\rho=\sum_{n=1}^N\rho_n,\qquad p=\sum_{n=1}^Np_n,\qquad p_n=w_n\rho_n,\quad w_n>-1,\quad n=1,2,\dots,N>1\; ,
\]
and the components do not interact. That means that the conservation equation in (\ref{goveq}) holds for every component separately, yielding $\rho_n=M_n/a^{3(w_n+1)}$, $M_n={\rm const>0}$, and thus  the analog of the governing problem (\ref{ydot}) becomes
\[
\dot{y}^2=y^{-\nu_1}+y^2-k\omega+\sum_{n=2}^N\,\mu_ny^{-\nu_n},\qquad y(0)=0\; ,
\]
with $y$, $\tau$ and $\omega$ normalized as in (\ref{var}), (\ref{par}) using $w_1$ and $M_1$, $\nu_n=3w_n+1$, and $\mu_n>0$ being the abundance of the $n$th component normalized appropriately. Writing $y(\tau,\nu_1,\nu_1,\dots,\nu_N, k\omega)$ for the solution, we have
\[
y(\tau,\nu_1,\nu_1,\dots,\nu_N, k\omega)>y(\tau,\nu_j, k\omega)
\]
for any pertinent $j$; generally, adding every new component enhances the expansion. Assuming $\nu_1$ is the smallest of all the powers, we see that at very large times the expansion is described by the one--component equation (\ref{ydot}) with $\nu=\nu_1$; however, the growth amplitude depends essentially on the whole expansion history, in other words, on all the equations of state. 

As for the other results, all the estimates of sec. 4 remain true, including inequalities (\ref{tauflat}) for the ages of open, flat, and closed universe (of course, the expression for $\tau_{Fl}$ depends now on all the parameters involved). The properly generalized transformation property (\ref{transol}) for the open universe solutions also holds; in fact, the number of such independent transformations  is equal to the number of matter components, $N$.

\section*{Acknowledgment}
This work was supported by NASA grant NAS 8-39225 to Gravity Probe B. I am grateful to R.J.Adler for initiating this study, and to him, A.D.Chernin, D.I.Santiago, and R.V.Wagoner for enlightening remarks. I thank the Gravity Probe B Theory Group for numerous discussions of the paper.

\section*{Appendix. Analysis of Power Series of $A_\nu(k\omega)$}

First we derive expression (\ref{Aser1}) for the derivatives $\phi^{(n)}_\nu(0)$. By (\ref{nderfi}), for $n=1,2,...$ we have:
\[
\frac{\sqrt{\pi}}{\Gamma(n+1/2)}\,\phi^{(n)}_\nu(0)=
\int_0^{\infty}\frac{x^{\nu(2n+1)/2}\,dx}{\left(1+x^{\nu+2}\right)^{n+1/2}}=
\frac{1}{\nu+2}\,
\int_0^{\infty}\frac{s^{\frac{\nu(2n+1)+2}{2(\nu+2)}-1}\,ds}{\left(1+s\right)^{n+1/2}}
\]
The last integral is one of the known representations of the beta--function $B(x,y)$ with the arguments
\[
x=\frac{\nu(2n+1)+2}{2(\nu+2)}=\frac{\nu n}{\nu+2}+\frac{1}{2},\qquad
y=n+\frac{1}{2}-x=\frac{2n}{\nu+2}\; ,
\]
which is easily verified by means of the substituion $s=t/(1-t)$. Therefore
\[
\frac{\sqrt{\pi}}{\Gamma\left(n+\frac{1}{2}\right)}\,\phi^{(n)}_\nu(0)=
\frac{1}{\nu+2}\,B\left(\frac{\nu n}{\nu+2}+\frac{1}{2},\,\frac{2n}{\nu+2}\right)=
\frac{1}{2n}\,\frac{2n}{\nu+2}\,
\frac
{\Gamma\left(\frac{\nu n}{\nu+2}+\frac{1}{2}\right)\Gamma\left(\frac{2n}{\nu+2}\right)}
{\Gamma\left(n+\frac{1}{2}\right)}=
\]
\[
\frac{1}{2n}\,\frac
{\Gamma\left(\frac{\nu n}{\nu+2}+\frac{1}{2}\right)\Gamma\left(\frac{2n}{\nu+2}+1\right)}
{\Gamma\left(n+\frac{1}{2}\right)}\; ,
\]
and this is, in fact, expression (\ref{Aser1}).

We now prove the asymtotic formula (\ref{coefas}) for the coefficients of the seires (\ref{Aser1}) in the general case. To do that, we use the large argument asymptoitcs of the gamma--function,
\[
\Gamma(z)=\sqrt{2\pi}\,\exp\left[(z-1/2)\ln z- z\right]\,\left[1+O(1/z)\right],\qquad
 z\to+\infty,
\]
and convenient notations
$$
u=\frac{2}{\nu+2},\qquad v=\frac{\nu}{\nu+2},\qquad u+v=1\;.
\]
We thus write:
\[
\frac{\phi^{(n)}_n(0)}{n!}=\frac{1}{2\sqrt{\pi}\,n}\,
\frac{
\Gamma\left(un+1\right)
\Gamma\left(vn+\frac{1}{2}\right)
}
{\Gamma\left({n}+1\right)}\; ,
\eqno(A.1)
$$
and then
$$
\frac{
\Gamma\left(un+1\right)
\Gamma\left(vn+\frac{1}{2}\right)
}{\Gamma\left({n}+1\right)}=
\sqrt{2\pi}\,\exp (Q_n)\left[1+O\left(\frac{1}{n}\right)\right]\qquad n\to+\infty\; , 
\eqno(A.2)
$$
where
\[
Q_n=(un+1/2)\,\ln(un+1)-(un+1)+(vn)\,\ln(vn+1)-(vn+1/2)-(n+1/2)\,\ln(n+1)+(n+1)=
\]
\[
(un+1/2)\,\ln(un)+(vn)\,\ln(vn)-(n+1/2)\,\ln n+O\,(1/n)=
\]
\[
n\,\ln(u^u v^v)+(1/2)\,\ln u+O\,(1/n)\;.
\]
Intorducing this into (A.2), and the resulting equality into (A.1), we end up with
$$
\frac{\phi^{(n)}_n(0)}{n!}=\frac{1}{2\sqrt{\pi}\,n}\,\sqrt{2\pi u}\,(u^u v^v)^n\,\left[1+O\left(\frac{1}{n}\right)\right]=
$$
$$
\frac{1}{\sqrt{\nu+2}}\,\frac{(u^u v^v)^n}{n}\,
\left[1+O\left(\frac{1}{n}\right)\right],\qquad n\to+\infty\; .\qquad\qquad
(A.3)
$$
It remains only to notice that
\[
u^uv^v=\left(\frac{2}{{\nu+2}}\right)^{2/(\nu+2)}\,
\left(\frac{\nu}{{\nu+2}}\right)^{\nu/(\nu+2)}=
\frac{2^{2/(\nu+2)}\,\nu\,^{\nu/(\nu+2)}}{\nu+2}=
\left(\frac{2}{{\nu}}\right)^{2/(\nu+2)}\frac{\nu}{\nu+2}=\frac{1}{\omega_\nu}\; ,
\]
in order to see that (A.3) is identical to (\ref{coefas}).
\vfill
\eject
\section*{References}
[1] A.G. Riess et al., {\it Astron. J.} 116, 1009 (1998).

\noindent[2] S. Perlmuter et al., {\it Astrophys. J.} 517, 565 (1999).

\noindent[3] A.G. Riess et al., astro-ph/0104455 (to appear in {\it Astrophys. J.}).

\noindent[4] R.R. Caldwell, R. Dave, P.J. Steinhard, {\it Phys.Rev.Lett.} 80, 1582 (1998).

\noindent[5] A.D.Chernin, D.I.Santiago, A.S.Silbergleit. astro-ph/0106144 (submitted to {\it Phys. Lett.A}).
 
\noindent[6] D.Kramer, H.Stephani, E.Herlt, M.MacCallum. Exact Solutions of Einstein's Field Equations. Ed. E.Schmutzer. Cambridge University Press, 1980.

\noindent[7] H.Bateman, A.Erd\'ely. Higher Transcendental Functions, vol. 3. McGraw-Hill Book Co., 1955.

\noindent[8] Ya.B.Zeldovich. Zeldovich JETP  41, 1609, 1961 (Rus)
\vfill\eject
\vskip5mm
\begin{figure}[htb]
\vskip3in\relax\centerline{\hbox to4in{\special{bmp:lamfig1.pdf x=4in,y=3in}\hfil}}
\caption{Parameters for the Closed Universe Solution}
\end{figure}
\vfill\eject
\vskip5mm
\begin{figure}[htb]
\vskip3in\relax\centerline{\hbox to4in{\special{bmp:lamfig2.pdf x=4in,y=3in}\hfil}}
\end{figure}
\vskip 5mm
\centerline{Figure 2. $A_\nu(km)$ for nonnegative $\nu$:}
\centerline{0 - $\nu=0$, 1 - $\nu=1$, 2 - $\nu=2$, AS - lower bound $0.5\sqrt{(-k\omega)}$}
\vfill\eject
\vskip5mm
\begin{figure}[htb]
\vskip3in\relax\centerline{\hbox to4in{\special{bmp:lamfig3.pdf x=4in,y=3in}\hfil}}
\vskip 5mm
\centerline{Figure 3. $A_\nu(km)$ for negative $\nu$:}
\centerline{1 - $\nu=-1/2$, 1 - $\nu=-1$, 3 - $\nu=-3/2$, AS - lower bound $0.5\sqrt{(-k\omega)}$}
\end{figure}
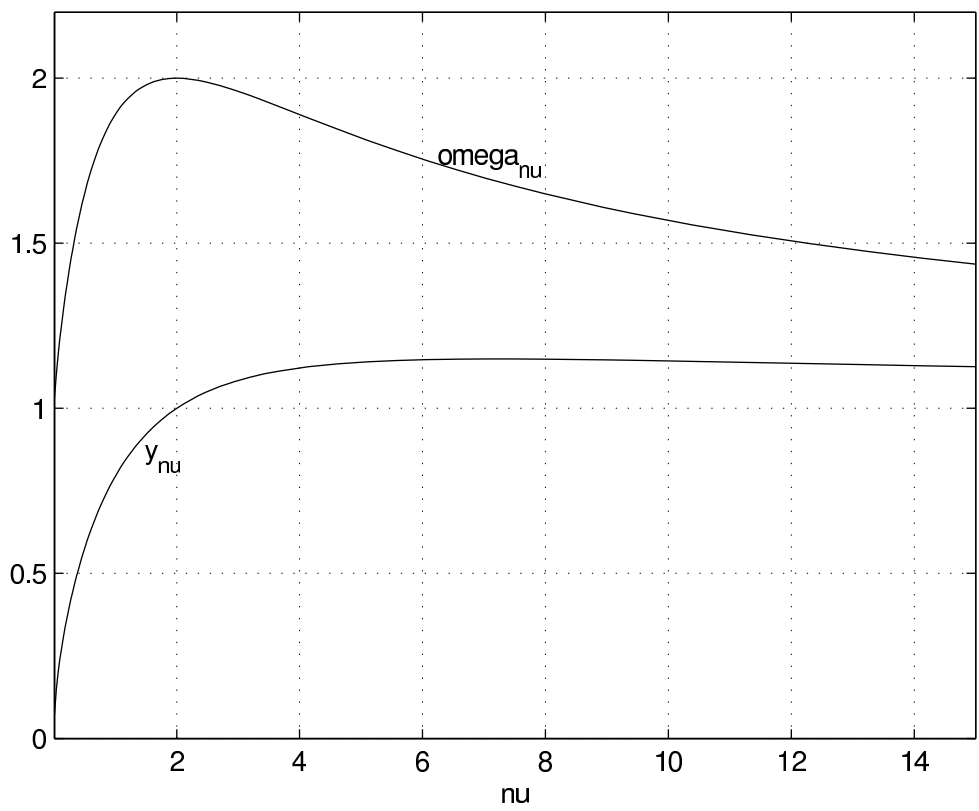
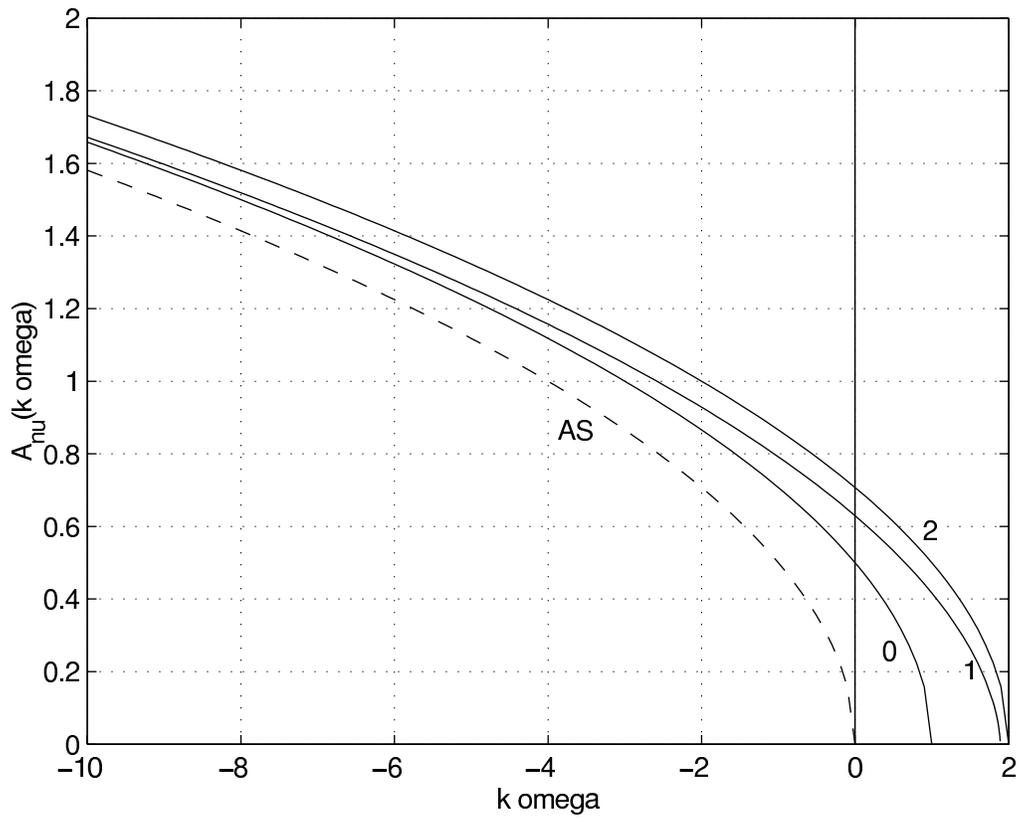
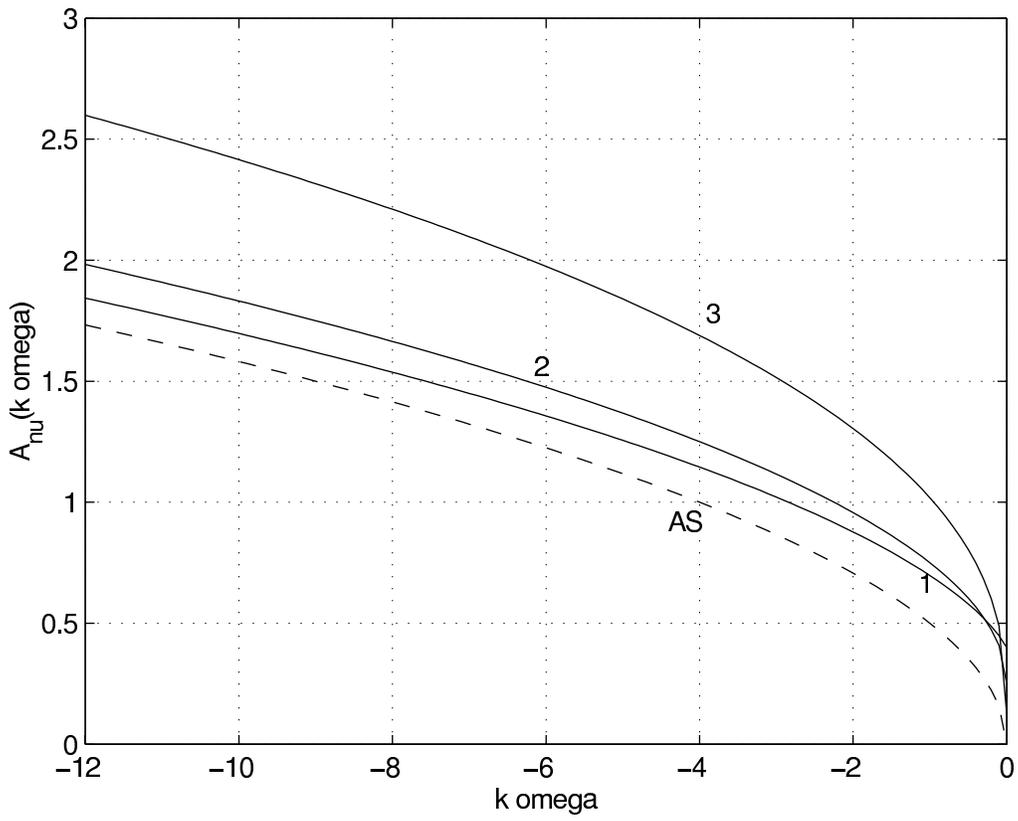
\end{document}